\def\ident{\equiv}

\def\pref#1{(\ref{#1})}   

\newcommand{\newc}{\newcommand}
\newc{\Agut}{A^{GUT}}
\newc{\Aew}{A^{EW}}
\newc{\Bgut}{B^{GUT}}
\newc{\Bew}{B^{EW}}
\newc{\gsim}{\lower.7ex\hbox{$\;\stackrel{\textstyle>}{\sim}\;$}}
\newc{\lsim}{\lower.7ex\hbox{$\;\stackrel{\textstyle<}{\sim}\;$}}
\newc{\ie}{{\it i.e.}}          
\newc{\etal}{{\it et al.}}
\newc{\eg}{{\it e.g.}}          
\newc{\kev}{\hbox{\rm\,keV}}            
\newc{\mev}{\hbox{\rm\,MeV}}            
\newc{\gev}{\hbox{\rm\,GeV}}            
\newc{\tev}{\hbox{\rm\,TeV}}
\newc{\ddet}{\hbox{\rm det}}
\newc{\ecm}{\,e\,\hbox{\rm cm}}

\newc{\mtop}{m_t}
\newc{\mbot}{m_b}
\newc{\mz}{m_Z}
\newc{\mw}{m_W}
\newc{\alphasmz}{\alpha_s(m_Z^2)}
\newc{\swsq}{\sin^2\theta_W}
\newc{\tw}{\tan\theta_W}
\newc{\cw}{\cos\theta_W}
\newc{\sw}{\sin\theta_W}
\newc{\BR}{\hbox{\rm BR}}
\newc{\xpb}{\hbox{\rm\, pb}}
\newc{\zbb}{Z\to b\bar}
\newc{\Gb}{\Gamma (Z\to b\bar b)}
\newc{\Gh}{\Gamma (Z\to \hbox{\rm hadrons})}
\newc{\rbsm}{R_b^\hbox{\rm sm}}
\newc{\rbsusy}{R_b^\hbox{\rm susy}}
\newc{\drb}{\delta R_b}

\newc{\tbeta}{\tan\beta}
\newc{\uL}{{\tilde u_L}}
\newc{\uR}{{\tilde u_R}}
\newc{\cL}{{\tilde c_L}}
\newc{\cR}{{\tilde c_R}}
\newc{\tL}{{\tilde t_L}}
\newc{\tR}{{\tilde t_R}}
\newc{\dL}{{\tilde d_L}}
\newc{\dR}{{\tilde d_R}}
\newc{\sL}{{\tilde s_L}}
\newc{\sR}{{\tilde s_R}}
\newc{\bL}{{\tilde b_L}}
\newc{\bR}{{\tilde b_R}}
\newc{\mhp}{m_{H^\pm}}
\newc{\mhalf}{m_{1/2}}
\def\NPB#1#2#3{Nucl. Phys. {\bf B#1}, #3 (19#2)}
\def\PLB#1#2#3{Phys. Lett. {\bf B#1}, #3 (19#2)}

\def\PRD#1#2#3{Phys. Rev. {\bf D#1}, #3 (19#2)}
\def\PRL#1#2#3{Phys. Rev. Lett. {\bf#1}, #3 (19#2)}

\def\ZPC#1#2#3{Zeit. f\"ur Physik {\bf C#1}, #3 (19#2)}

\def\beq{\begin{equation}}
\def\eeq{\end{equation}}
\def\bea{\begin{eqnarray}}
\def\eea{\end{eqnarray}}
\def\slashchar#1{\setbox0=\hbox{$#1$}           
   \dimen0=\wd0                                 
   \setbox1=\hbox{/} \dimen1=\wd1               
   \ifdim\dimen0>\dimen1                        
      \rlap{\hbox to \dimen0{\hfil/\hfil}}      
      #1                                        
   \else                                        
      \rlap{\hbox to \dimen1{\hfil$#1$\hfil}}   
      /                                         
   \fi}                                         %
\catcode`@=11
\long\def\@caption#1[#2]#3{\par\addcontentsline{\csname
  ext@#1\endcsname}{#1}{\protect\numberline{\csname
  the#1\endcsname}{\ignorespaces #2}}\begingroup
    \small
    \@parboxrestore
    \@makecaption{\csname fnum@#1\endcsname}{\ignorespaces #3}\par
  \endgroup}
\catcode`@=12

\def\jfig#1#2#3{
 \begin{figure}
 \centering
 \epsfysize=2.75in
 \hspace*{0in}
 \epsffile{#2}
 \caption{#3}
 \label{#1}
 \end{figure}}
\def\bfig#1#2#3{
 \begin{figure}
 \centering
 \epsfysize=3.5in
 \hspace*{0in}
 \epsffile{#2}
 \caption{#3}
 \label{#1}
 \end{figure}}

\def\bmulteq{\begin{eqnarray}}
\def\emulteq{\end{eqnarray}}
\def\im{\bar}
\def\Im{{\rm Im}}
\def\Re{{\rm Re}}
\def\Arg{{\rm Arg}}
\def\abs#1{\left| #1 \right|}

\documentstyle[aps,preprint,epsf]{revtex}
\begin{document}
\draft

\begin{titlepage}
\preprint{
\noindent
\begin{minipage}[t]{3in}
\baselineskip=16pt
\begin{flushright}
SLAC-PUB-7286\\
hep-ph/9609511 \\
September 1996
\end{flushright}
\end{minipage}
}

\title{Constraints on supersymmetric soft phases from
renormalization group relations}

\author{Robert Garisto\footnote{
\baselineskip=16pt
Work supported in part by
the Department of Energy, contract DE-AC02-76CH00016.}}
\address{BNL Theory Group, Bldg 510a, Brookhaven National Lab, Upton, NY 11973}
\author{James D.~Wells\footnote{
\baselineskip=16pt
Work supported by 
the Department of Energy, contract DE-AC03-76SF00515.}}
\address{Stanford Linear Accelerator Center, Stanford University, 
Stanford, CA 94309}
\maketitle
%
%
\baselineskip=16pt
\begin{abstract}
\baselineskip=16pt
By using relations derived from renormalization group equations (RGEs),
we find that strong indirect constraints can be placed on 
the top squark mixing phase in $A_t$ from the electric dipole 
moment of the neutron ($d_n$).
Since $m_t$ is large, any GUT-scale phase in
$A_t$ feeds into other weak scale phases through RGEs, which 
in turn contribute to 
$d_n$.  Thus $CP$-violating effects due to a weak-scale 
$A_t$ are strongly constrained.  We find that $|\Im A_t^{EW}|$ must
be smaller than or of order $|\Im B^{EW}|$, making the electric dipole
moment of the top quark unobservably small in most models.
Quantitative estimates of the contributions to $d_n$ from $A_u$, $A_d$ and $B$
show that substantial fine-tuning
is still required to satisfy the experimental bound on $d_n$.
While the low energy phases of the $A$'s are not as strongly constrained as
the phase of $B^{EW}$, we note that
the phase of a universal $A^{GUT}$ induces large contributions
in the phase of $B^{EW}$ through RGEs, and is thus still strongly 
constrained in most models with squark masses below a TeV.
\end{abstract}
\end{titlepage}


\baselineskip=18pt

\section{Introduction}

Supersymmetry (SUSY)~\cite{review} is
one of the most compelling extensions of the Standard Model.   
It is the only known perturbative solution to the naturalness
problem~\cite{Ibanez}, it unifies the gauge coupling constants
for the observed value of $\sin^2\theta_W$, it allows
radiative EW symmetry breaking,
and the lightest SUSY partner provides a good dark matter candidate.
SUSY models with such features are generally in
excellent agreement with experiment, and there is even the possibility
that a recent  CDF event \cite{CDF event} is of supersymmetric origin
\cite{CDF interp}.    

One of the few phenomenological problems associated with SUSY models
is their generically large predictions for the electric dipole moment (EDM)
of the neutron, $d_n$.  
Supersymmetric models with universal soft breaking parameters have two physical
phases, beyond the CKM and strong phases of the SM,
which can be taken to be the triscalar and biscalar soft breaking parameters
$A$ and $B$.  
These phases give a large contribution to 
$d_n$, of order $10^{-22} (100\gev/M_{susy})^2 e\, cm$,
where $M_{susy}$ is a characteristic superpartner mass.
The experimental
upper bound on $d_n$ is of order $10^{-25} e\, cm$ \cite{dnexpt}, so that if
superpartner masses are near the weak scale, the phases of these complex
soft parameters must be fine-tuned to be less than or of order
$10^{-2}$--$10^{-3}$ since there is no {\it a priori} reason for
them to be small \cite{first susycp}.
If one wants to avoid 
such a fine-tuning, there are two approaches:
suppress $d_n$ with very large squark masses (greater than a TeV)
\cite{kizukuri oshimo}, or
construct models in which the new SUSY phases naturally vanish
\cite{susy CP sols}.  Models with very heavy squarks are unappealing because
in such models LSP annihilation is usually suppressed enough so that the
relic density is unacceptably large \cite{falk olive srednicki}.  
They also lead to a fine-tuning
problem of their own in getting the $Z$ boson mass to come out right in
EW symmetry breaking.

It is natural to consider solutions of the second type, and demand that
the soft phases are zero by some symmetry.
While that would leave only a small CKM
contribution to 
$d_n$~\cite{dugan85:413,gerard-kurimoto,SUSY CKM dn,bertolini vissani},
and thus avoid any fine-tuning in meeting the experimental bound on $d_n$,
it would also mean that there is no non-SM $CP$ violation,
which is needed by most
schemes for electroweak baryogenesis~\cite{bgenesis}. Also,
such models do not generate signals of non-SM $CP$ violation,
such as those involving top squark mixing.  
There are ways of naturally obtaining small nonzero soft phases which
leave sufficient $CP$ violation for baryogenesis
\cite{garisto modCP,smallphase,dimo95:220,gutcp}, but these phases would
still have to meet the bounds from $d_n$ and would probably be unobservably
small in most EW processes---unless the soft terms are not universal.

Recently it has been pointed out that large non-SM
$CP$-violating top quark couplings
could be probed at high energy colliders~\cite{dtrefs}.
A measurement of 
a large top quark EDM, for example, would indicate physics beyond the SM,
and it is interesting 
to ask whether SUSY models can yield an observable effect.
Several references have attempted to use $CP$ violation from top squark
mixing due to the complex parameter $A_t$ to yield large $CP$-violating
effects in collider processes involving top quarks \cite{Atrefs}.
Such papers either explicitly or implicitly assume nonuniversal
soft couplings $A_q$ at the GUT scale; otherwise, the phase
of $A_t$ would be trivially constrained by $d_n$.  We consider
whether it is possible to obtain
large effects due to the phase of
$A_t$ at the EW scale by relaxing the universality of $A$.
We will show that due to renormalization group induced 
effects on other low energy phases,
the phase of $A_t$ is strongly constrained by $d_n$, and it is not
possible, for most areas of parameter space, to have large $CP$-violating
effects due to the imaginary part of $A_t$.

We will assume that no parameters are fine-tuned and thus we will require
the phases 
at the GUT scale to be either identically 
zero (presumably through some symmetry)
or no less than $1/10$.  If one permits an arbitrary degree of fine-tuning, the
whole SUSY $CP$ violation 
issue becomes moot, and one can derive no constraints on the phase of $A_t$.  
While one can construct models which give small universal phases, as we said
above, the 
fine-tuning needed to evade the constraints we derive is unlikely to be 
explained naturally.
Our approach in this paper is to assume the reasonable fine-tuning 
criterion we have just outlined,
and ask what it implies about low energy SUSY $CP$-violating phenomenology.

In Sec. II, we review the basics of SUSY $CP$ violation.  We present our
results derived from RGEs in Sec. III, and impose the
neutron EDM constraints on $\Im A_t$ using those results in Sec. IV.
In Sec. V we discuss top squark mixing induced $CP$ violating observables
in more detail in light of our constraints on the phase of $A_t$, and
we give some concluding remarks in Sec. VI.  The details
from Sec. III are written up in Appendix A, and the full one-loop
calculation for the SUSY contribution to the neutron EDM is given 
in Appendix B.


\section{SUSY $CP$-violating phases}

The soft breaking potential in the MSSM is
\bmulteq
-{\cal L}_{soft} & = & {1 \over 2} \abs{m_i}^2 \abs{\varphi_i}^2 + 
{1 \over2} \sum_\lambda M_{\lambda} \lambda \lambda  +
\nonumber\\
 & &
\epsilon_{ij}[A_U  \tilde U^*_R Y_U \tilde Q_L^i] H_u^j +
\epsilon_{ij}[A_D \tilde D^*_R Y_D^\dagger \tilde Q_L^j ] H_d^i +
\epsilon_{ij}[A_E \tilde E^*_R Y_E^\dagger \tilde L_L^j] H_d^i  + \\
& & \epsilon_{ij} B \mu H_u^i H_d^j  + h.c. \nonumber\\
\label{eq:Lsoft}
\emulteq
where we take $A_U = \hbox{\rm diag}\{A_u,A_c,A_t\}$,
$A_D = \hbox{\rm diag}\{A_d,A_s,A_b\}$, $A_E = 
\hbox{\rm diag}\{A_e,A_\mu,A_\tau\}$;
$Y_U$, $Y_D$, and $Y_E$ are the Yukawa coupling matrices;
$\tilde Q$, $\tilde L$, $\tilde U_R$, $\tilde D_R$ and
$\tilde E_R$ are the squark and slepton fields;
$\lambda$ are the gauginos and $\varphi_i$ are the scalars
in the theory.

A common simplifying assumption is that this soft Lagrangian arises as
the result of a GUT-scale supergravity (SUGRA) model with universal
soft triscalar coupling $A$, gaugino mass $M_\lambda = M_{1/2}$, and scalar
mass $m_i=m_0$. This provides
an explanation for the absence of flavor changing neutral currents 
which arise from loops with squarks of nondegenerate mass~\cite{FCNC}. 
Such supersymmetric models have only two independent physical 
$CP$-violating phases
beyond the CKM and strong phases of the SM~\cite{dugan85:413} although these
phases appear in several different linear combinations in low
energy phenomenology \cite{dimo95:220,garisto natCP}. 
We will take the two physical phases to be $\Arg A$ and $\Arg B$.  

It turns out that all $CP$ violating vertices in this model 
arise through the diagonalization
of complex mass matrices~\cite{garisto modCP}.  The complex quantities which
appear in these matrices are $A_q + \mu^* R_q$ and $\mu$, where
$R_q$ is $\tan\beta$ (the ratio of Higgs vacuum expectation values)
for $q=d,\ s,\ b$ and $\cot\beta$ for $q=u,\ c,\ t$,
and where the phase of $\mu$ is simply equal to the phase of $B^*$ by 
a redefinition of fields.
Thus for $d_n$,
which involves only $u$ and $d$ quarks, there are only contributions from
three low energy combinations of the two SUSY GUT 
phases: $\Arg(A_d- \mu\tan\beta)$, 
$\Arg(A_u- \mu\cot\beta)$,
and $\Arg\mu$. (In the Appendix B, a complete expression of $d_n$ is given which
includes suppressed contributions from phases of the other squark mixings.)

Even with universal boundary conditions, the elements of the 
matrices $A_U$, $A_D$ and $A_E$ 
have distinct phases at the EW scale because of renormalization group evolution.
We will also relax, in some places, the assumption that their phases
started the same at the GUT scale.
We assume (for simplicity) that these
matrices are diagonal.  One possible consequence of this approach is that
one could have $d_n\simeq 0$ because $\Im A_d$ and $\Im A_u\simeq 0$,
but other $A_q$, notably $A_t$, could have large phases which
lead to observable effects.  
These include angular correlations and polarizations \cite{Atrefs},
including effects attributable to
the electric dipole moment of the top quark, $d_t$.
As discussed in the Introduction, this scenario is strongly constrained
by RGE running.


\section{Renormalization group flow of complex soft terms}

The goal of this section is to demonstrate how a large phase
in $A_t$ can feed into other parameters in the theory through
renormalization group running.
The imaginary part of $A_t$ 
at the weak scale, $\im A_t^{EW}$, is determined 
by running $\im A_t^{GUT}$ (and for large $\tan\beta$, $\im A_b^{GUT}$)
down to the weak scale via the renormalization group equations (RGEs).  
(For compactness of notation, we will define $\im x=\Im x$ in the following
sections.)
We will show that large $\im \Aew_t$ induces potentially large
values of $\im B^{EW}$ and $\im \Aew_{u,d}$, which give an unacceptably
large neutron electric dipole moment.

Rather than write RGEs for the whole effective theory, we need only 
consider a complete subset of them which includes $A_q$ and $B$.
The running of these soft terms depends upon the gaugino masses,
the top and bottom Yukawas 
(we ignore tiny effects from the other Yukawa couplings)
and the gauge coupling constants $\alpha_a= \lambda_a^2/4\pi$ 
($a=1,\ 2,\ 3$).
We define $t= 1/4 \pi \ln(Q/M_{GUT})$ and write

\bea
{d M_a \over d t} & = & 2 b_a \alpha_a M_a \\
{d A_t \over d t} & = & 2 c_a \alpha_a M_a + 12 \alpha_t A_t + 2 \alpha_b A_b 
\label{At RGE}\\
{d A_{u,c} \over d t} & =  & 2 c_a \alpha_a M_a + 6\alpha_t A_t 
\label{Auc RGE} \\
{d A_b \over d t} & = & 
2 c_a' \alpha_a M_a + 2 \alpha_t A_t + 12 \alpha_b A_b \\
{d A_{d,s} \over d t} & = & 2 c_a' \alpha_a M_a + 6 \alpha_b A_b \\
{d B \over d t} & = & 2 c_a''' \alpha_a M_a + 6 \alpha_t A_t + 6 \alpha_b A_b \\
{d \alpha_t\over dt} & = & 2\alpha_t\left( -c_a\alpha_a +6\alpha_t
                           +\alpha_b\right) 
\label{alphat full RGE}\\
{d\alpha_b\over dt} & = & 
2\alpha_b\left( -c'_a\alpha_a+\alpha_t+6\alpha_b\right) \\
{d \alpha_a\over dt} & = & 2b_a\alpha_a^2 
\label{alphaa RGE}
\eea
where $a$ is summed from 1 to 3, and
\bea
        b_a & = & \left( {33\over 5},1,-3 \right), \\
        c_a & = & \left( {13\over 15},3,{16\over 3} \right), \\
        c_a' & = & \left( {7\over 15},3,{16\over 3}\right), \\
        c_a''' & = & \left( {3\over 5},3,0\right),
\eea
and the Yukawa coupling constants $\alpha_{t,b} = \lambda_{t,b}^2/4\pi$
are related to the masses by
\beq
\lambda_t  =  \frac{g_2}{\sqrt{2}}\frac{m_t}{m_W}\frac{1}{\sin\beta}, \
\lambda_b  =  \frac{g_2}{\sqrt{2}}\frac{m_b}{m_W}\frac{1}{\cos\beta}.
\eeq
We note that some references \cite{attwo} list the 
$\alpha_t A_t$ coefficient in \pref{Auc RGE} as 2, but we have 
confidence that the coefficient is actually 6 \cite{bertolini vissani,atsix}.
Nevertheless our conclusions do not depend qualitatively on this coefficient.

We are mainly interested in
the evolution of $\im A_q$ and $\im B$.  We can set the phase of
the common gaugino mass to zero at the GUT scale by a phase rotation and
then $\im M_i =0$ at all scales.  Therefore the RGE for the imaginary
parts of the $A_q$ and $B$ can be written without the $M_a$ terms:
\bea
{d \im A_t \over d t} & = &  12 \alpha_t \im A_t + 2 \alpha_b \im A_b, 
\label{ImAt RGE}\\
{d \im A_b \over d t}  & = & 2 \alpha_t \im A_t + 12 \alpha_b \im A_b, 
\label{ImAb RGE}\\
{d \im A_{u}\over dt} & = & 6\alpha_t \im A_t, \\
{d \im A_{d}\over dt} & = & 6 \alpha_b \im A_b, \\
{d \im B\over dt} & = & 6\alpha_t\im A_t + 6\alpha_b\im A_b .
\eea
Using the above RGEs, we can derive the following useful relations:
\bea
\Delta \im B \ & = & \Delta \im A_{u,c} + \Delta \im A_{d,s} =
  {6\over 14} \left( \Delta \im A_t + \Delta \im A_b \right),\\
\Delta \im A_{u,c} & = & {3 \over 35} 
  \left(6 \Delta \im A_t - \Delta \im A_b \right),\\
\Delta \im A_{d,s} & = & {3 \over 35}
  \left(6 \Delta \im A_b - \Delta \im A_t \right),
\eea
where $\Delta \im B = \im B^{GUT} - \im B^{EW}$, etc.  For small $\tan\beta$,
we can neglect $m_b$ so that these relations simplify to
\bea
\Delta \im B\ & = &
 \Delta \im A_{u,c} = 3 \Delta \im A_b = {1\over2} \Delta\im A_t,
\nonumber
 \\
\Delta \im A_{d,s} & = & 
0.
\label{Delta eqs no mb}
\eea
Thus, given the GUT values, 
to obtain the low energy values for the imaginary parts of all the 
soft terms, one only needs to find $\im A_t^{EW}$ and $\im A_b^{EW}$,
and for small $\tan\beta$, we only need the former.

In the small $\tan\beta$ limit ($\alpha_b\simeq 0$), we can use 
Eq. \pref{ImAt RGE} to obtain the ratio of EW to GUT scale values
of the imaginary part of $A_t$:
\beq
r_t \ident \im A_t^{EW}  / \im A_t^{GUT}  = 
\exp\left[-\int_{t_{EW}}^{t_{GUT}} 12 \alpha_t(t) dt \right] .
\label{def rt}
\eeq
If the top quark were light, the integral in Eq. \pref{def rt} would be
small and $r_t$ would be close to one, but since the top quark is heavy,
we find that $r_t$ is well below one.
We can use
the relations in Eq. \pref{Delta eqs no mb} 
and the definition for
$r_t$ in \pref{def rt} to relate the low energy values for the imaginary
parts of $A_t$ to $B$ and $A_u$ (for small $\tan\beta$):

\beq
\im A_t^{EW} = {-2 r_t \over 1- r_t} \left( \im B^{EW} - \im B^{GUT} \right)= 
{-2 r_t \over 1- r_t} \left( \im A_u^{EW} - \im A_u^{GUT} \right).
\label{At intof B}
\eeq
We will make the simplifying assumption that $\im A_u^{GUT}$ and 
$\im B^{GUT}$ are zero.  As we will see in the next section, this is
reasonably well justified by our fine-tuning criterion, at least
for the phase of $B$. 

Next, we must find $r_t$.
We obtain a pseudo-analytic
solution to Eq. \pref{def rt} in terms of EW and GUT scale quantities 
in Eq. \pref{AppA no mb solution} of Appendix A, but this is useful only 
if one has already obtained
the GUT values for the $\alpha$'s by numerical integration of the
RGEs.  
While we cannot find a truly analytic
solution to Eq. \pref{def rt}, we can place an analytic 
upper bound on $r_t$ which is 
sufficient to make our point.  
We note that the integral in Eq. \pref{def rt} is simply the area under
the curve of the top Yukawa $\alpha_t$ as it runs from the EW scale to
the GUT scale. Thus we can
place an upper bound on $r_t$ simply by finding a lower bound to that
area.  
In Appendix A, we do this by placing a lower bound on $\alpha_t(t)$ at each $t$,
and we obtain 
\beq
r_t \lsim 1 - 12 \alpha_t^{EW}/f_{EW},
\label{no mb bound rt}
\eeq
which is valid for small $\tan\beta$ so long as $12 \alpha_t^{EW} < f_{EW}$.
Here $f_{EW}\ident 2 c_a \alpha_a^{EW}\simeq 1.5 + 32/3(\alpha_s^{EW} - .12)$,
so, for example, Eq. \pref{no mb bound rt} is valid for $m_t=175$ 
if $1.3<\tan\beta\ll m_t/m_b$
(for smaller $\tan\beta$, $r_t$ gets closer to zero, but does not actually
reach it).
Thus we have placed an analytic bound on the running of $A_t$ completely
in terms of EW quantities.  For $\alpha_s(M_Z)=.12$, $\sin\beta\rightarrow 1$
(moderate $\tan\beta$) and $m_t=175$ ($m_t=160$), we find 
that $r_t<.43$ ($r_t<.52$), which, from Eq. \pref{At intof B}, corresponds to
$|\im A_t^{EW}| < 1.5 |\im B^{EW}|$  
($|\im A_t^{EW}| < 2.2 |\im B^{EW}|$).  For small $\tan\beta$, the bound is
even stronger, so that for $\tan\beta$ small enough to neglect $m_b$ effects,
we obtain
\beq
|\im A_t^{EW}| < 2.2\,\hbox{\rm min}\,
\left\{ |\im B^{EW}|,|\im A_u^{EW}|\right\}
\eeq
and in practice the coefficient is less than 2.

In Fig.~1, we plot $r_t$ ($= \im A_t^{EW}/\im A_t^{GUT}$) as a function 
of the top Yukawa coupling for different values of $\alpha_s(M_Z)$ in
the limit where effects proportional to $m_b$ can be ignored.  For 
$m_t> 160$ GeV, $\lambda_t$ is always greater than about $0.87$ 
for all values of $\tan\beta$, which
means that $r_t$ is always
less than $.45$, in agreement with our analytic bounds.
Also plotted are $-\im B^{EW}/\im A_t^{GUT}= (1-r_t)/2$, and
$-\im B^{EW}/\im A_t^{EW}= (1-r_t)/2r_t$, which is greater than 
1 ($0.6$) for $m_t=175$ ($160$).  Thus $|\im A_t^{EW}| \lsim  |\im B^{EW}|$, in
agreement with our analytic results.

Next we consider moderate $\tan\beta$,
where one must take into account
the mixing of $\im A_t$ and $\im A_b$ but where $\tan\beta$ is not of
order $m_t/m_b$. For $\im A_b^{GUT}/\im A_t^{GUT} > 0$,
these effects lower $r_t$, and one can simply use the $m_b=0$ upper bound 
on $r_t$ 
derived above.\footnote{There is a subtlety for the  case of small positive
$\im A_b^{GUT}/\im A_t^{GUT}$ for which there 
can be a net positive contribution to
$r_t$ if  $\im A_b$ runs down below zero.  However, the maximum effect
on the bound is very small.
}

For $\im A_b^{GUT}/\im A_t^{GUT} < 0$ (recall that with universal $A$ this 
ratio would simply be $+1$), one simply maximizes the positive contribution to
$r_t$ from $\im A_b$ to obtain (see Appendix A)
\beq
r_t < 1 - 12 \alpha_t^{EW}/f_{EW} -
{1\over 6} 
\left(\im A_b^{GUT} / \im A_t^{GUT}\right)
{\alpha_b^{EW} \over \alpha_t^{EW} - \alpha_b^{EW}} .
\label{mb bound rt}
\eeq
Note that the last term raises the upper bound on $r_t$, but the effect is
small until $\tan\beta$ gets quite close to $m_t/m_b$.
For $m_t=175\gev$, $\im A_b^{GUT} = -\im A_t^{GUT}$, and 
$\tan\beta = .7 m_t/m_b \simeq 35$ (recall that we are evaluating
all quantities at the EW scale, so $m_b$ is somewhat lower than 
the value at $q^2=m_b^2$), we find the bound $r_t < 0.6$.

Effects due to $m_b$ are evident in Figs.~2-4, which show $\im A_t^{EW}$,
$\im B^{EW}$, $\im A_u^{EW}$ and $\im A_d^{EW}$, 
normalized to $\im A_t^{GUT}$, as
a function of $\tan\beta$ for various GUT-scale boundary conditions.
In Fig.~2, only the phase of $A_t^{GUT}$ is non-zero, while in Figs.
3 and 4, $\im A_b^{GUT}$ has values of $+\im A_t^{GUT}$ and $-\im A_t^{GUT}$
respectively.  In all cases, $r_t$ (the solid curve) remains below $0.35$ and
has its largest value just below $\tan\beta=m_t/m_b$ for 
$\im A_b^{GUT}/\im A_t^{GUT}<0$
(Fig.~4), in agreement with our analytic results.  
This means that the EW value for the phase of $A_t$ is constrained to 
be less than about a third, {\it independent} of constraints from
low energy $CP$ violating observables.
The magnitude of the 
imaginary part induced into $\im B^{EW}/\im A_t^{GUT}$ by $\im A_t$ is greater
than $0.35$ except for large $\tan\beta$ and $\im A_b^{GUT}/\im A_t^{GUT}<0$.  
For $\tan\beta=m_t/m_b$ and $\im A_b^{GUT}/\im A_t^{GUT}<0$, 
$\im B^{EW}$ actually
goes through zero, because $\Delta \im B$ gets
equal and opposite contributions from $\Delta \im A_t$ and $\Delta \im A_b$
there.  At that point the ``t'' and ``b'' RGE coefficients are almost exactly
the same at each $t$ (because the Yukawa coupling runnings differ only in a 
small U$(1)$ coefficient), and the boundary conditions have opposite
signs, so that $\im A_t(t) \simeq - \im A_b(t)$ for all $t$.
Of course $\im A_u^{EW}$ and $\im A_d^{EW}$ are non-zero because they involve
different linear combinations of $\Delta \im A_t$ and $\Delta \im A_b$,
so there is still a strong constraint on $\im A_t$ from $d_n$ there.

Finally we note that for large $\tan\beta$, one can place constraints on
$\im A_b^{GUT}$ as well, since it can then affect other low energy phases 
through renormalization group running.  For $\tan\beta \sim m_t/m_b$, the
constraints are of the same order as on $\im A_t^{GUT}$, while for small
$\tan\beta$, $\im A_b^{GUT}$ is unconstrained
(though the $\im A_t$ contribution to $\im A_b^{EW}$ for small $\tan\beta$
is constrained  to be small and $\im A_b^{EW}$ 
can be large only if $|\im A_b^{GUT}| \gg |\im A_t^{GUT}|$).


\section{Bounds from the neutron EDM}

Now that we have placed an upper bound on the magnitude of $\im A_t$
in terms of $\im A_u$, $\im A_d$, and $\im B$, we need to explore the
constraints on the latter three imaginary parts (in low energy observables,
we will drop the label EW).  As we mentioned in 
the Introduction,
one of the strongest constraints on
$CP$ violating phases is the electric dipole moment (EDM) of the
neutron, $d_n$. In Appendix B, we write expressions for the full
supersymmetric contribution to $d_n$.  One sees that all the pieces are
proportional to $\im A_u$, $\im A_d$, or $\im \mu$ (except for the negligibly
small pieces proportional to $\im A_q$).  We can redefine the
Higgs fields so that
the phase of $\mu$ is just the opposite of the phase of $B$, and
thus
\beq
\im \mu = - \left| {\mu \over B}\right| \im B 
= \left| {\mu \over B}\right| 
\left( {1- r_t \over 2 r_t} \im A_t - \im B^{GUT} \right),
\label{mu intof At}
\eeq
where the RHS follows for small $\tan\beta$.

In order to estimate the size of $\im \mu$ we will need
an estimate of $|\mu/B|$ in Eq. \pref{mu intof At}.
We can find this ratio by considering the two equations which $\mu$ and
$B$ need to satisfy to ensure that EW symmetry breaking occurs and that
the $Z$ boson gets the right mass:
\bea
{2B\mu}&=&-(m^2_{H_u}+m^2_{H_d}+2\mu^2)\sin 2\beta, 
\label{Bmu}
\\
\mu^2&=&-\frac{m^2_Z}{2}+\frac{m^2_{H_d}-m^2_{H_u}\tan^2 \beta}{\tan^2\beta-1}.
\label{musq}
\eea
In the limit that $\tan\beta \to \infty$, we see that the right hand side
of Eq.~\pref{Bmu} goes to zero, so $B\to 0$, whereas
$\mu^2$ is not forced to zero.  For $\tan\beta \to 1$, the right hand
side of Eq.~\pref{musq} blows up forcing $\mu$ to take on very large
values.  When $\mu^2$ dominates Eq.~\pref{Bmu} and $\tan\beta =1$ then
we are led to a value of $|B|=|\mu|$.  
So in both the $\tan\beta \to \infty$ limit and the $\tan\beta \to 1$ limit we 
find that $|\mu|\geq |B|$.  We have run thousands of
models numerically~\cite{kkrw} which include the one-loop corrections to 
Eqs.~\pref{Bmu} and~\pref{musq} and found that $|\mu| \gsim |B|$ is indeed
a good relationship for most of the parameter space.  As expected,
it is violated most strongly for intermediate values of $\tan\beta$.
For example, for $\tan\beta =10$ we have found a small region of
parameter space where $|\mu|/|B|$ is as low as 0.4, although most solutions
prefer $|\mu|/|B|>1$.  
We will assume that $|\mu|/|B| \gsim 1$, and thus the 
fine-tuning constraint on the phase of $B$ is even
stronger than on what we obtain below for the phase of $\mu$.

{}From Appendix B, we see that $d_n$ can be written in terms of the three
imaginary parts,
\beq
{d_n \over 10^{-25} e \, cm}
=k_n^{A_u} {\im A_u \over m_0} + k_n^{A_d} {\im A_d \over m_0}
+ k_n^{\mu} {\im\mu \over m_0}
=k_n^{A_u} {\im A_u \over m_0} + k_n^{A_d} {\im A_d \over m_0}
- k_n^{\mu} \left|{\mu\over B}\right| {\im B \over m_0},
\label{dn u d mu}
\eeq
where we have normalized the RHS by the SUSY mass scale $m_0$, and the LHS
by the region of the experimental bound so that
the coefficients $k$ are dimensionless.  
We can rewrite the EW imaginary parts in Eq. \pref{dn u d mu} using 
Eq. \pref{Delta eqs no mb} as
\beq
{d_n \over 10^{-25} e \, cm} = {d_n^{GUT} \over 10^{-25} e \, cm} +
{1 - r_t \over 2 r_t} 
\left( - k_n^{A_u} + k_n^{\mu} \left| {\mu \over B}\right| \right) 
{\im A_t \over m_0} ,
\label{dn t}
\eeq
where $d_n^{GUT}/10^{-25} e \, cm$ is just Eq. \pref{dn u d mu} with EW 
values of $\im A_{u,d}$ and $\im B$ 
replaced by GUT quantities.  It vanishes if $\im A_{u,d}^{GUT}$
and $\im B^{GUT}$ are zero.
In supergravity models,
$|A^{GUT}|$ and $|B^{GUT}|$ are
of order $m_0$, so that barring fine-tuned cancellations, the GUT scale
phases must be less than order $1/k_n$.  If the $k$'s are greater than
order 10, then our fine-tuning criterion dictates that we set the GUT
phases to zero (presumably protected by some symmetry).  Thus we need
an estimate of the $k_n's$.

In Figs. 5a, 5b, and 5c, we plot the values for 
$k^{A_u}_n$, $k^{A_d}_n$, and $k^\mu_n$ respectively
in many different models as a function of squark mass, and as a function of
$\tan\beta$ in Figs. 5d, 5e, and 5f.
We see that $k^{A_u}_n$
and $k^{A_d}_n$ are fairly flat functions of $\tan\beta$, whereas
$-k^\mu_n$ increases with $\tan\beta$ due to the $\mu\tan\beta$
terms in the expression for $d_n$.  We also see that most models give
$k^{A_u}_n > 2$ $(0.8)$, $k^{A_u}_n > 7$ $(3)$,
and $|k^\mu_n| > 100 (40)$, for squark masses below
500 GeV (1 TeV),
so that order one phases in all the SUSY complex quantities usually give 
a neutron EDM which is of order 100 (40) times the experimental bound.
We note that these are substantially larger contributions (and thus
stronger constraints) than claimed by the recent work of Falk and Olive
\cite{falk olive constraints}, though this is probably due to the 
fact that they use very heavy squark masses in an effort to find the
smallest fine-tuning of phases consistent with cosmology.
While one can argue whether or not the bounds on the phases of $A_{u,d}$ 
represent a fine-tuning, the bound on the 
phase of $\mu$ (and thus $\im B^{EW}$, which comes from $\im B^{GUT}$
and $\im A_t^{GUT}$) certainly does.  Thus,
by our fine-tuning criterion, the phases of $B^{GUT}$ and 
$A_t^{GUT}$ should be zero.  We note
that in the case of $universal$ $A$ it is $irrelevant$ whether or
not the low energy phases of $A_u$ and $A_d$ are strongly constrained, since
the phase of the universal $A^{GUT}$ makes a large contribution to the
low energy value of $\im \mu$  (since 
$\im A_t^{GUT} = \im A^{GUT}$).

To give an idea of what level of neutron EDM one expects with different initial
assumptions, we plot in Fig.~6
$d_n/10^{-25} e \,$ cm with universal $|A^{GUT}|$ for three cases:
(a,d) $\Arg A_t^{GUT} = \Arg A_b^{GUT}=0.1$ 
and all other phases zero,   
(b,e) $\Arg A_t^{GUT} = -\Arg A_b^{GUT}=0.1$
and all other phases zero,    
and (c,f) universal 
phases $\Arg A^{GUT} = \Arg B^{GUT} = 0.1$.  As one can see, 
even with phases of order $0.1$, most models
have an absolute value for $d_n/10^{-25} e \, cm$ greater than one, 
inconsistent with the experimental bounds.

As can be gathered by the spread of points in the scatter plots and the 
number of parameters involved, the results depend on one's
model assumptions.  For example, if one requires $\tan\beta$ to be 
small (say because of $b$-$\tau$ unification), and the squarks are
allowed to be very heavy, then there is very little fine-tuning
needed for the current experimental bound on $d_n$.  On the other hand, 
if SUSY is detected at LEP 2 or TeV 33, then 
even the smallest $\tan\beta$ models would require fine-tuning.

In minimal supergravity models the natural scale for the $A$ terms is $m_0$.  
In Fig.~7 we have plotted $d_n/10^{-25}e\cdot$cm versus
$Im A_t^{EW}/m_0$ to succinctly demonstrate how quickly 
the EDM rises when $Im A^{EW}_t\neq 0$.  To construct this plot
we chose a random phase for $A_t$ at the GUT scale, 
forced all other
phases equal to zero at that scale, and then ran all the
parameters down to the weak scale.  
A sharp drop in $d_n$ occurs at $Im A^{EW}_t/m_0\simeq 0$ because 
$\Im A_t^{GUT}$ can be small there and thus induces only small phases into the
other low energy soft parameters.  Models with $d_n$ around $10^{-25}e\cdot$cm
at $Im A^{EW}_t/m_0 \simeq 0$ occur for low $\tan\beta$ where 
$\Im A_t^{GUT} \gg \Im A_t^{EW}$ but where $d_n$ otherwise tends to be smaller.
This means that most models with $d_n$ below the experimental bound in 
Fig.~6 also have a small EW value for $\Im A_t$, and thus from
Fig.~7 we can place a
stronger constraint on $\Im A_t^{EW}$ than we obtained on
$\Im A_t^{GUT}$: $Im A^{EW}_t/m_0\lsim 1/20$.

Thus we conclude that models with 
universal GUT-scale phases of the soft parameters, and models in which only
$A_t^{GUT}$ has a non-zero phase,
have difficulty meeting the bounds from $d_n$ and our
fine-tuning criterion.  Models with non-zero $\im A_{c}^{GUT}$, 
$\im A_{s}^{GUT}$, or $\im A_{l}^{GUT}$ can meet the constraint from $d_n$
without fine-tuning, as can those with non-zero $\im A_{b}^{GUT}$ for
small $\tan\beta$.  For the remainder of the paper, we will for simplicity
set all the
GUT-scale phases to zero except for that of $A_t$.
Even though our fine-tuning criterion implies  that $\im A_{t}^{GUT}$
should be zero, we find it useful
to ask what effects one would have if one allows that fine-tuning.


\section{The top quark EDM}

Now that the top quark has finally been discovered, one can envision
some nice experiments which measure properties of this known particle.
Future colliders, such as the NLC, can provide many precision measurements
of the production cross-section and decay properties of the top quark.
It is possible that signatures of new physics could arise out of such
a study.  One property of the top quark which has received 
much attention~\cite{dtrefs} is the possibility of measuring its EDM by looking
at the decay distributions of the $t\bar t$ pairs. (Other CP-violating 
observables are possible, such as those arising from $t\to bW$ decays,
but we will make our point only with the top EDM.) It is generally
estimated that the top quark EDM ($d_t$) can be measured to values as low as
${\cal O}(10^{-18})\,e\,\hbox{cm}$~\cite{dtrefs}.  Given the 
constraints which we derived above, we ask if the minimal supersymmetric
standard model can yield a value for $d_t$ this large.

In the context of supersymmetry, it has been proposed~\cite{Atrefs}
that a large $d_t$  
is possible if the phase of $\Aew_t$ is of order one.
But in Sec. III,  we showed that $\im\Aew_t$ is constrained to be smaller than
or of order the phases which contribute to $d_n$.  The EDM of the top
is thus constrained to be less than a constant times the neutron EDM:
\beq
\frac{d_t}{d_n} \lsim \xi \frac{m_t}{m_d}
\frac{\hbox{\rm det}M^2_{\tilde q}}{\hbox{\rm det}M^2_{\tilde t}},
\label{dtdn}
\eeq
where $\hbox{\rm det}M^2_{\tilde q}=m^2_{\tilde q_1}m^2_{\tilde q_2}$
is the determinant of the (down) squark mass-squared matrix, and the value
of $\xi$ depends upon many different SUSY parameters, but is generically
of order 1.
Normalizing $d_n$ to the experimental bound, we see that
\beq
d_t \lsim 
\xi
\frac{\hbox{\rm det}M^2_{\tilde q}}{\hbox{\rm det}M^2_{\tilde t}}
{d_n^{expt} \over 10^{-25}\,e\,\hbox{cm} }
\, 2 \times 10^{-21}\,e\,\hbox{cm}.
\label{dteq}
\eeq
In addition to this constraint, we recall that the phase of $A_t$ at the
EW scale must be less than about $1/3$, just from the RGE suppression
factor $r_t$.  Thus, as long as $\ddet M_{\tilde d}\simeq \ddet M_{\tilde t}$,
we expect $d_t$ to fall about three orders of magnitude below
detectability at proposed future high energy colliders.  

We can turn this analysis around.  If a large top quark EDM is
discovered, can it be explained in the MSSM?
One possibility is that
a conspiracy occurs between {\it several} large phases in the
theory to render $d_n$ below experimental limits, yet produce
a $d_t$ detectable at high energy colliders.  This is equivalent
to saying that all the ${\cal O}(1)$ coefficients which
we absorbed into the parameter $\xi$ in Eq.~\pref{dteq} actually
conspire to give $\xi \gsim 10^{3}$.  As we argued in
the Introduction, we would not view this as a likely explanation.

Another possibility to consider is that the 
top squarks are
much lighter than the other squarks.  For $d_t$ to be observable, we
would need the determinants in Eq. \pref{dtdn}
to have a ratio $\gsim 10^3$.
This is possible, but it too would require some fine-tuning.
The large top-quark-induced running of the $\tilde t_R$ goes in
the right direction---the lightest top squark mass eigenvalue tends to
be smaller than the 
other quarks.  However, $\tilde t_2$ generally tracks fairly well
with the other squarks, $\tilde q_L$, and  thus, we estimate that
\beq
\frac{\ddet M^2_{\tilde d}}{\ddet M^2_{\tilde t}}\lsim
\frac{m^2_{\tilde d}}{m^2_{\tilde t_1}},
\eeq
which means that we would need $m_{\tilde t_1}\lsim 
m_{\tilde d}/\sqrt{1000}$ to yield an observable $d_t$.
If experiment determines that 
$m_{\tilde t_1}>80\gev$ then this condition would imply that the
superpartners of the light quarks are above $2.5\tev$.  
This is essentially the heavy squark ``solution" to the $CP$ violation
problem we mentioned in the Introduction, with an additional 
fine-tuning implied by the small ratio $m_{\tilde t_1}/m_{\tilde d}$.

Finally, one could appeal to differences between $d_t$ and $d_d$ due
to effects proportional to $m_t^2/v^2$, which are negligible in $d_d$.
To achieve $\xi$ of order $10^3$, one again needs a fine-tuned conspiracy
of couplings.  

Thus we conclude that if a large $d_t$ were found, one would probably
have to look beyond the MSSM for an explanation.


\section{Concluding remarks}

It has long been noted that the phases of soft supersymmetric parameters
generically lead to an unacceptably large neutron EDM.  This fine-tuning
problem has slowly become less vexing as the theoretical expectations
for the squark masses have risen faster than the experimental bound
on the neutron EDM has fallen.  Nevertheless, for squark masses below
about a TeV, we showed in Sec. V that the phase of $B$ and universal phase
of $A$ do not meet the fine-tuning criterion set forth in the Introduction
(see Fig 6c).  Certainly,  if supersymmetry is discovered at LEP 2 or TeV 33, 
a fundamental explanation for the absence of a neutron EDM would be needed,
and any scheme for baryogenesis at the EW scale would require that
mechanism to leave small effective low energy phases in the soft terms
\cite{garisto modCP,smallphase,dimo95:220,gutcp}. 

{}From the phenomenological point of
view, it is tempting to postulate that the soft phases are not universal---that
the EW phase of $A_t$ is large, while the
other phases which directly contribute to the neutron EDM are small.
This would allow interesting signatures of supersymmetric
$CP$ violation to be visible in top quark physics at future colliders.   
But we have demonstrated by using the renormalization group equations that 
the imaginary part of $A_t$ must be less than twice the imaginary part of
$B$, and $A_t$-induced CP-violating observables such as the top EDM are 
thus expected to be unobservably small in almost all minimal SUSY models.

These constraints are particularly important for models of EW baryogenesis
which rely upon the phase of the stop LR mixing parameter,
$A_t - \mu \tan\beta$, to generate enough $CP$ violation for baryogenesis.
Such models must also have sufficiently small $|A_t - \mu \tan\beta|$
to ensure that the phase transition is first order~\cite{riotto}.
There has also been a recent attempt to explain the observed $CP$ violation
in the neutral kaon system with zero CKM phase and non-zero off-diagonal
phases in the general $A$ matrices \cite{abel frere}. 
If the universal diagonal $A$ parameter has a large phase at the GUT scale, it
will, as we noted above, give a large contribution to $d_n$ through a
renormalization group induced phase in $\mu$, as well as from a direct 
contribution.  One could evade such bounds by insisting that the off-diagonal
components of the $A$ matrices have a large phase, while the phases of the
diagonal $A$'s and of $B$ vanish. Although this hypothesis can probably be
technically consistent with
our fine-tuning criterion (phases either zero or large), this
scenario strikes us as unnatural.
\medskip

\noindent
{\it Acknowledgements.} We would like to thank G.~Kane,
S.~P.~Martin, D.~Wyler, S.~Thomas,
A.~Riotto, and A.~Soni
for helpful discussions.  RG greatly appreciates the hospitality of
the Brookhaven National Lab HEP Theory Group.


\section{Appendix A}

In this appendix, we provide the details related to our analytic results
of Sec. III.  It is interesting to note that we can use the RGEs
for the top Yukawa and gauge coupling constants in Eqs. \pref{alphat full RGE} 
and \pref{alphaa RGE} to write a pseudo-analytic solution to
$r_t = \Im A_t^{EW} / \Im A_t^{GUT}$.  The integral in Eq. \pref{def rt}
can be rewritten as
$\ln(\alpha_t^{EW}/\alpha_t^{GUT}) -
\sum_a (c_a/b_a) \ln(\alpha_a^{GUT}/\alpha_a^{EW})$, which allows us to
write a pseudo-analytic $r_t$ in terms of EW and GUT scale quantities
(the latter of which cannot be found analytically):
\beq
r_t = {\alpha_t^{EW} \over \alpha_t^{GUT}}
\Pi_{a=1}^3 \left( {\alpha_a^{EW} \over \alpha_a^{GUT}} \right)^{c_a/b_a}.
\label{AppA no mb solution}
\eeq

To place an analytic upper bound on $r_t$, we must place a lower bound
on the area $\int_{t_{EW}}^{t_{GUT}} 12 \alpha_t(t) dt$.  We will need
the $\alpha_b=0$ limit of the running of the top Yukawa coupling in 
\pref{alphat full RGE},
\beq
{d \alpha_t \over d t} = -f(t) \alpha_t + 12 \alpha_t^2,
\label{no mb alphat RGE}
\eeq
where $f(t) = 2 c_a \alpha_a$. While this cannot be solved analytically,
we note that $\alpha_3(t)$ runs down with energy and one can show 
that $f(t)$ will 
be at its maximum value at the EW scale. Thus if we take $f(t)$ to 
the constant $f_{EW}$, we will minimize the running of $\alpha_t$, and
Eq \pref{no mb alphat RGE} can be solved analytically to yield the
bound 
\beq
\alpha_t(t) > \left( {12 \over f_{EW} } + 
\left( {1 \over \alpha_t^{EW} } - {12 \over f_{EW}} \right) e^{f_{EW}(t-t_{EW})}
\right)^{-1},
\label{bound alphat}
\eeq
which is valid for $12 \alpha_t^{EW} < f_{EW}$ (larger $\alpha_t^{EW}$ allows
the bound on $\alpha_t(t)$ to reach infinity for $t < t_{GUT}$ and thus  
makes the bound useless), which corresponds to $\tan\beta > 1.3$
for $m_t=175$.
If we replace $\alpha_t(t)$ in the integral above by the RHS of 
Eq. \pref{bound alphat},
we can find an analytic solution for the lower bound on the area
which, for the relevant range of
$f_{EW}$ and $t_{EW}$, can be approximated by
\beq
-f_{EW} t_{EW} - \ln\left[{12 \alpha_t^{EW} \over f_{EW}} + 
\left(1 - {12 \alpha_t^{EW} \over f_{EW}}\right) e^{-f_{EW} t_{EW}} \right] 
\simeq
- \ln \left[ 1 - {12 \alpha_t^{EW} \over f_{EW} } \right],
\eeq
which yields Eq. \pref{no mb bound rt} directly.

For moderate $\tan\beta$, we need to include $m_b$ effects
which mix $\im A_t$ with $\im A_b$, and $\alpha_t$ with $\alpha_b$.
The coupled differential equations \pref{ImAt RGE} and \pref{ImAb RGE} 
can be solved
analytically only 
if the coefficients, which here are proportional to $\alpha_t$ and
$\alpha_b$, are constants.  To obtain bounds on the running of $\im A_t$ and
$\im A_b$, we can break up the range of energy from $t_{EW}$ to $t_{GUT}$
into small regions where the coefficients are effectively constant, and
iteratively evolve from the GUT scale down to the weak scale. At each 
energy $t_j$, the value for $\im A_t$ is given by

\bea
\im A_t\left(t_{j+1}\right) &&\simeq \im A_t(t_j) 
\exp\left(-12 \alpha_t(t_j) \delta t \right)
\nonumber \\
&&- {1 \over 6}\im A_b(t_j) 
\left\{
{\alpha_b(t_j) \over \alpha_t(t_j) - \alpha_b(t_j)}
\left[
\exp\left(-12 \alpha_b(t_j) \delta t\right) - 
\exp\left(-12 \alpha_t(t_j) \delta t\right)
\right] \right\},
\label{ImAt iteration eq}
\eea
provided that $T\ident \alpha_b/(\alpha_t - \alpha_b)$ is not large.  Here 
$\delta t = t_j - t_{j+1}$, which is positive.  Iterating 
Eq. \pref{ImAt iteration eq}
gives a complicated expression with terms proportional to each of the
$T(t_j)$'s.  However, each of these terms is positive,
so that taking $T(t_j)$ to its maximum value maximizes the size of the
quantity in $\{\ \}$'s in Eq. \pref{ImAt iteration eq}, which is what
we need for the case $\im A_b^{GUT}/\im A_t^{GUT} <0$.  Once we take 
$T(t_j)\rightarrow T_{max}$, many terms cancel, and we are left with 
(taking $\delta t\rightarrow 0$)  the upper limit

\bea
\im A_t^{EW}&& < \im A_t^{GUT} 
\exp\left(-\int_{t_{EW}}^{t^{GUT}} 12 \alpha_t(t) dt \right)
\nonumber\\
&&- {1 \over 6}\im A_b^{GUT} 
\left({\alpha_b \over \alpha_t - \alpha_b}\right)_{max}
\left[
\exp\left(-\int_{t_{EW}}^{t^{GUT}} 12 \alpha_b(t) dt \right)
-
\exp\left(-\int_{t_{EW}}^{t^{GUT}} 12 \alpha_t(t) dt \right)
\right].
\label{ImAt upper bound intermed}
\eea
One can show analytically that $T(t)$ reaches its maximum value at the lowest
energy of the range, and thus we can replace $T_{max}$ by 
$\alpha_b^{EW}/(\alpha_t^{EW} - \alpha_b^{EW})$.  To obtain a simpler
bound, one can reduce the 
$[\ ]$'s in Eq.  \pref{ImAt upper bound intermed} to $1$ by taking 
a lower bound on $\alpha_b(t)$ to be zero and an upper bound on 
$\alpha_t(t)$ to be infinity.  Finally one uses the $m_b\simeq0$ 
bound on $r_t$ obtained in
Eq. \pref{no mb bound rt} 
for the first term in Eq. \pref{ImAt upper bound intermed} to
obtain the upper bound on $r_t$ in Eq. \pref{mb bound rt}.


\section{Appendix B}

In this Appendix we present analytic expressions for 
the full one-loop SUSY contribution to
the neutron electric dipole moment, $d_n$.   The gluino \cite{gerard-kurimoto}
and chargino \cite{kizukuri oshimo} contributions appear in the literature.
While  an expression for the neutralino contribution is given by
Kizukuri and Oshimo \cite{kizukuri oshimo}, it is written in terms of
$4\times 4$ complex unitary matrices which must be determined numerically.
Below we give an expression for this neutralino
contribution solely in terms of the
mass matrices (and other MSSM parameters), and a useful approximation to that
expression, which do not require calculating complex unitary matrices.

To find the neutron EDM, we first calculate the EDM of the up and
down quarks ($d_q$) from one loop diagrams with photons attached to either
($a$) an internal boson or ($b$) an 
internal fermion line.  Then the neutron EDM
is related to the quark EDM's in the Naive Quark Model by
$d_n =(4d_d - d_u)/3$, though recent work has
argued that this expression overestimates $d_n$ if the strange
quark carries a large fraction of the neutron and proton spin \cite{Ellis EMC}.
The Feynman integrals associated with ($a$) and ($b$) are \cite{Garisto Kl3}:
\beq
I^a(x) = {1 \over (1-x)^2} \left[ -{3 \over 2} + {x \over 2} 
- {\ln x \over 1-x} \right], \
I^b(x) = {1 \over (1-x)^2} \left[ {1 \over 2} + {x \over 2} 
+ {x \ln x \over 1-x} \right] .
\label{Feynman Is}
\eeq

As we mentioned earlier, all SUSY $CP$ violating effects arise from
diagonalizing complex mass matrices \cite{garisto modCP}.
Gluino loops contribute to the quark EDM $d_q$ through the complex phase in the
left-right mixing elements for up and down squarks:  
\beq
 d_q(\tilde g) = {-2 \over 3 \pi} Q_q e \alpha_s
{m_q m_{\tilde g} {\rm Im}(A_q - \mu R_q) \over 
m_{\tilde q_0}^4}
I^b \Bigl({m_{\tilde g}^2 \over m_{\tilde q_0}^2} \Bigr) .
\label{dq gluino}
\eeq
We have averaged over the nearly degenerate squark mass eigenstates for 
simplicity: $m_{\tilde q_0}^2 = m_{\tilde q_1} m_{\tilde q_2}$ and 
$I(x_0)= (I(x_1)+I(x_2))/2$.  Here $Q_q e$ is the charge of quark $q$,
$R_q = \tan\beta \ (\cot\beta)$ for $q=d$ ($u$), and $m_{\tilde g}$
is the gluino mass.  (Note that we use $\Im z$ for the
imaginary part of $z$ in this appendix
because it is clearer than $\im z$ in more complicated expressions.)

The chargino contribution is proportional to the imaginary part of products
of elements of the matrices $U$ and $V$ which diagonalize the chargino
mass matrix.  It turns out that one can write those products
directly in terms of the elements of the chargino mass matrix, so that
the chargino contribution to $d_q$ can be written

\bea
d_q(\tilde\chi^+) &=& {e\over (4 \pi)^2} \Biggl\{
g^2  R_q  
{m_q m_{\tilde W} \Im\mu
\over 
m_{\tilde q'_0}^4 } 
{[\omega I^a(y_1) + Q_{q'} I^b(y_1)] - 
[\omega I^a(y_2) + Q_{q'} I^b(y_2)] \over y_1 - y_2 }   + \nonumber\\
&&{m_q \Re\mu \over \sin\beta \cos\beta} [\omega I^a(y_0) + Q_{q'} I^b(y_0)]
\sum_{r=q'_i}^{dsb\ {\rm or}\ uct} 
{\Im (A_r - \mu R_{q'}) \over  m_{\tilde r_0}^4}
{m_r^2 |V_{qr}|^2 \over v^2}
\Biggr\} ,
\label{dq chargino}
\eea
where $m_{\tilde W}$ is the wino mass, $\omega =+1$ ($-1$) for $q=d$ ($u$),
$y_1 = m_{\tilde\chi_1}^2/m_{\tilde q_0'}^2$, and $I(y_0)= (I(y_1)+I(y_2))/2$.
The primed quantities refer to the SU(2) partner, so if $q=d$, then $q'=u$ and
$r$ is summed over the set $\{u,c,t\}$.
Previous expressions for $d_q(\tilde\chi^+)$ have neglected the squark mixing
piece, which is the second term in Eq. \pref{dq chargino}. This piece
is suppressed relative to the other contributions 
by $m_r^2 |V_{qr}|^2 / v^2$, which is less than $10^{-4}$ for
$q=u$ or $d$ (but it can affect the EDM's of other quarks), 
though it is interesting that there is a (tiny) $direct$
contribution to $d_n$ from $\Im A_t$.

The neutralino contribution,
\bea
d_q(\tilde\chi^0) &&= {-Q_q e\over (4 \pi)^2} {1 \over m_{\tilde q_0}^2}
{m_q \over v_q}  \Biggl\{
\sum_{h=1}^{1,2} (a_{Lh}^q - a_{Rh}^q) \Im\Phi_{h\hat q} + \nonumber\\
{\Re(A_q + \mu R_q) v_q \over  m_{\tilde q_0}^2}
\sum_{h,l}^{1,2}&&a_{Lh}^q a_{Rl}^q \Im\Phi_{hl} -
{\Im( A_q - \mu R_q) v_q \over  m_{\tilde q_0}^2}
\sum_{h,l}^{1,2} a_{Lh}^q a_{Rl}^q
\Re\Phi_{hl}
\Biggr\},
\label{dq neutralino}
\eea
arises from the 4x4 complex neutralino mass matrix.  The index 
$\hat q = 3$ ($4$) for $q=d$ ($u$).
Recall \cite{gunion haber} that the ``1" and ``2" weak eigenstates are gauginos,
and the ``3" and ``4" weak eigenstates are higgsinos which couple to down and
up quarks respectively.  Thus ``34" and ``43" terms are absent, which will
allow us to simplify expressions involving the neutralino mass matrix,
since that is the position of the complex coefficient $\mu$. We have dropped
terms of order $m_f^2/v^2$ relative to the others.
The gauge coefficients $a_{Li}$ are:
\bea
a^q_{L1} &=& \sqrt{2} g \tan\theta_w \left( Q_q - T^q_{3L} \right)
= \sqrt{2} g \tan\theta_w /6 ,
\\
a^q_{L2} &=& \sqrt{2} g T^q_{3L},
\eea
and the  $a_{Ri}$ are the same as the 
$a_{Li}$ with $T_{3L} \rightarrow T_{3R} =0$.
The neutralino phases appear through a 4x4 matrix

\beq
\Phi_{hl} = \sum_{i=1}^4 U^T_{hi} \hat M_{ii} I^b_{ii} U_{il},
\label{def Phi}
\eeq

\noindent
where $U$ diagonalizes the neutralino mass matrix $M$, and $\hat M$  is the
diagonal result.  Here $I^b_{ij} = I^b(x_i) \delta_{ij}$, $i.e.$ $I^b_{ij}$
is the diagonal
matrix of Feynman integrals for the corresponding mass eigenvalues in
$\hat M$.  In the limit that the $I^b(x_i)$ are equal, the 
real part of $\Phi_{hl}$ can simply be written

\beq
\Re \Phi_{hl} \simeq I^b(x_0) \Re M_{hl},
\label{Re Phi}
\eeq
where $I^b(x_0) = \sum_i^4 I^b(x_i)/4$.
The imaginary part of $\Phi_{hl}$ is more difficult because it vanishes
in the limit of degenerate neutralino masses (except for the irrelevant
``34" and ``43" terms).  We know that $\Im\mu$ is the only complex coefficient
in the neutralino mass matrix $M$, so we can write

\beq
\Im \Phi_{hl} \simeq \Omega_{hl} \Im\mu,
\label{def Omega}
\eeq
where $\Omega_{hl}$ is a real matrix to be determined.
This allows us to see that $d_q(\tilde\chi^0)$ is proportional to 
$\Im\mu$ and $\Im(A_q - \mu R_q)$, just as the other contributions are.
It turns out that
$\Im \Phi_{hl}$ is proportional to $\Im(MM^*M)_{hl}$, $\Im(MM^*MM^*M)_{hl}$,
and $\Im(MM^*MM^*MM^*M)_{hl}$ (except for the ``34" and ``43" pieces).
To extract the $\Im\mu$ dependence, we ignore all terms of higher
order in $\Im\mu/|\mu|$, which is a valid approximation for the phases
allowed by the experimental bound on $d_n$.
Then these products (for $(h,l) \neq (3,4), \ (4,3)$)
simplify as follows: 
\bea
&\Im(MM^*M)_{hl} &\simeq \Im\mu \sum_{p=0}^2 (-1)^p 
(M_R^p P M_R^{2-p})_{hl},
\\
&\Im(MM^*MM^*M)_{hl} &\simeq \Im\mu \sum_{p=0}^4 (-1)^p 
(M_R^p P M_R^{4-p})_{hl},
\\
&\Im(MM^*MM^*MM^*M)_{hl} &\simeq \Im\mu \sum_{p=0}^6 (-1)^p 
(M_R^p P M_R^{6-p})_{hl},
\eea

\noindent
where $M_R=\Re M$ and $P$ is a matrix with $-1$ in the 34 and 43
positions and 0 everywhere else (so that $\Im(\mu P) = \Im M$).
After some calculation, we obtain an expression for the imaginary part of
the complex matrix $\Phi$:

\bea
\Im\Phi_{hl} \simeq
\Omega_{hl} \Im\mu \simeq
 {3 \over 2}\Im\mu \sum_{s=1}^4 
\Bigl(I(x_s) - \sum_j^{\neq s} I(x_j)\Bigr) \times 
&&\nonumber\\
\Biggl[\sum_{i,j,k}^{\neq s} \epsilon_{ijk} \hat M_i^4 \hat M_j^6
\sum_{p=0}^2 (-1)^p (M_R^p P M_R^{2-p})_{hl}                 - &&
\sum_{i,j,k}^{\neq s} \epsilon_{ijk} \hat M_i^2 \hat M_j^6
\sum_{p=0}^4 (-1)^p (M_R^p P M_R^{4-p})_{hl}                 +
\nonumber \\
&&\sum_{i,j,k}^{\neq s} \epsilon_{ijk} \hat M_i^2 \hat M_j^4
\sum_{p=0}^6 (-1)^p (M_R^p P M_R^{6-p})_{hl} \Biggr]   \times 
\label{ImPhi exact}\\
\Biggl[\sum_{i,j,k}^{\neq s} \epsilon_{ijk} \hat M_i^4 \hat M_j^6
\Bigl( 3 \hat M_s^2 - \sum_n^{\neq s} \hat M_n^2 \Bigr)           - && 
\sum_{i,j,k}^{\neq s} \epsilon_{ijk} \hat M_i^2 \hat M_j^6
\Bigl( 3 \hat M_s^4 - \sum_n^{\neq s} \hat M_n^4 \Bigr)           +
\nonumber\\
&&\sum_{i,j,k}^{\neq s} \epsilon_{ijk} \hat M_i^2 \hat M_j^4
\Bigl( 3 \hat M_s^6 - \sum_n^{\neq s} \hat M_n^6 \Bigr)  \Biggr]^{-1} ,
\nonumber
\eea

\noindent
where $\sum_j^{\neq s}$ means sum over the three members of the
set $\{1,2,3,4\} - \{s\}$. 
Here $\hat M_j$ are the mass eigenvalues of $\hat M$ ($i.e.$ the four physical
neutralino masses).

The expression above is completely analytic and exact except for the
approximation we made in dropping higher order terms in
$\Im\mu/|\mu|$, but it has so many terms that
it is not that useful.  Let us find an approximation to this expression
using the information about the neutralino mass eigenstates, namely
that they are fairly close together and the heaviest
neutralino is lighter than the squarks ($x_4\ll 1$) in most SUSY models.
This means that we can take
a simple linear fit to the Feynman integral by evaluating
$I^b(x)$ at the lowest and highest values of $x$:
\beq
 I^b(x) \simeq   K_0 + K_1 x \simeq I^b(x_1) + S_{41} (x-x_1) ,
\label{I linear}
\eeq

\noindent
where $S_{41}$ is the slope 
\beq
 S_{41} = {I^b(x_4) - I^b(x_1) \over x_4 - x_1 },
\eeq
and $x_j = m_{\tilde\chi_j^0}/m_{\tilde q_0}$.
Thus $K_1 = S_{41}$ and $K_0 =  I^b(x_1) - S_{41} x_1$. Note that
this approximation gives $exact$ values for $x_1$ and $x_4$, and is only
off for $x_2$ and $x_3$---a rough estimate is that the approximation is
correct to about $5\%$.

If we plug Eq. \pref{I linear} into ${\rm Im}\Phi_{hl}$ in Eq. \pref{def Phi}, 
we see that the
$K_0$ piece vanishes (except for the ``34" and ``43" pieces),
and we obtain
\beq 
\Im\Phi_{hl} 
\simeq S_{41} {\Im (M M^* M)_{hl} \over m_{\tilde q0}^2}
\simeq {S_{41} \over m_{\tilde q0}^2} \Im \mu
\sum_{p=0}^2 (-1)^p
(M_R^p P M_R^{2-p})_{hl}. 
\label{ImPhi approx}
\eeq
The neutralino contribution to $d_q$ is found by plugging Eq. \pref{Re Phi}
for $\Re\Phi_{hl}$ and Eq. \pref{ImPhi exact} or Eq. \pref{ImPhi approx} 
for $\Im\Phi_{hl}$
into \pref{dq neutralino}.

Finally, we want to relate the expressions for the three SUSY contributions
to the quark EDM in Eqs. \pref{dq gluino}, \pref{dq chargino}, and
\pref{dq neutralino}
in terms of the coefficients $k_n$ from Section IV.
Using the Naive Quark Model, $d_n=4/3 d_d - 1/3 d_u$, 
we can write $k_n^x = 4/3 k_d^x - 1/3 k_u^x$ and
\beq
k_q^x =
{d_q^x \over 10^{-25} e\, cm} {m_0 \over \Im x} ,
\eeq
where $x = A_u$, $A_d$, or $\mu$, and $d_q^x$ is the contribution to 
$d_q$ from complex quantity $x$.
We can see that if we neglect the tiny second term of $d_q(\tilde\chi^+)$
in Eq. \pref{dq chargino}, then $k_n^{A_u}$ ($k_n^{A_d}$)
gets contributions only
from the gluino and neutralino contributions to $d_u$ ($d_d$), whereas
$k_n^{\mu}$ gets contributions from all three of the SUSY contributions 
to $d_u$ and $d_d$.


\jfig{phasefuncs}{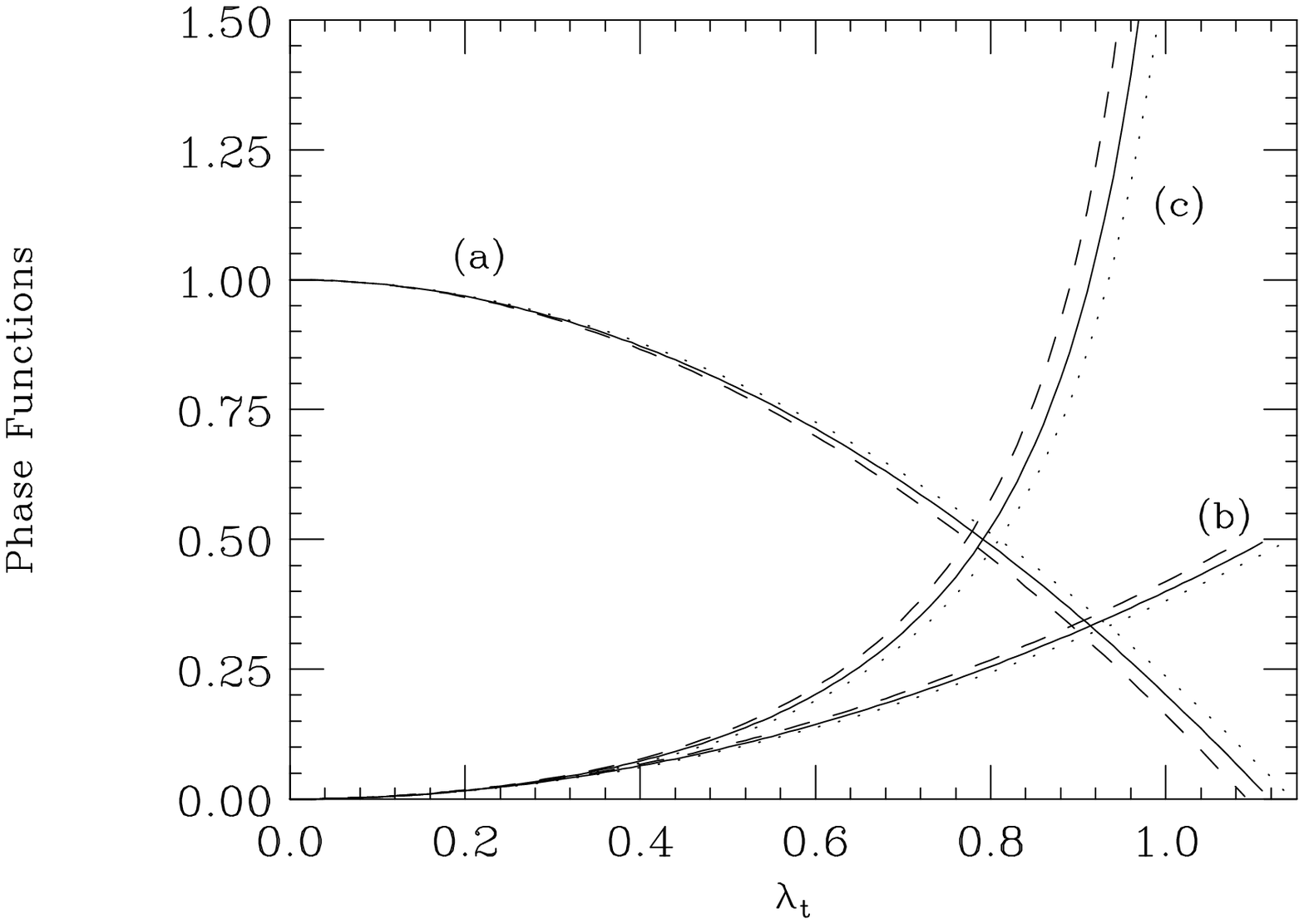}{Plot 
of the ratios (a) $r_t= \im A_t^{EW}/\im A_t^{GUT}$,
(b) $(1 -r _t)/2 = - \im B^{EW}/\im A_t^{GUT}$, and 
(c) $(1 -r _t)/2r_t = - \im B^{EW}/\im A_t^{EW}$ 
versus the top quark Yukawa coupling
for $\alpha_s(M_Z)=0.118\pm 0.006$.}

\jfig{atfig}{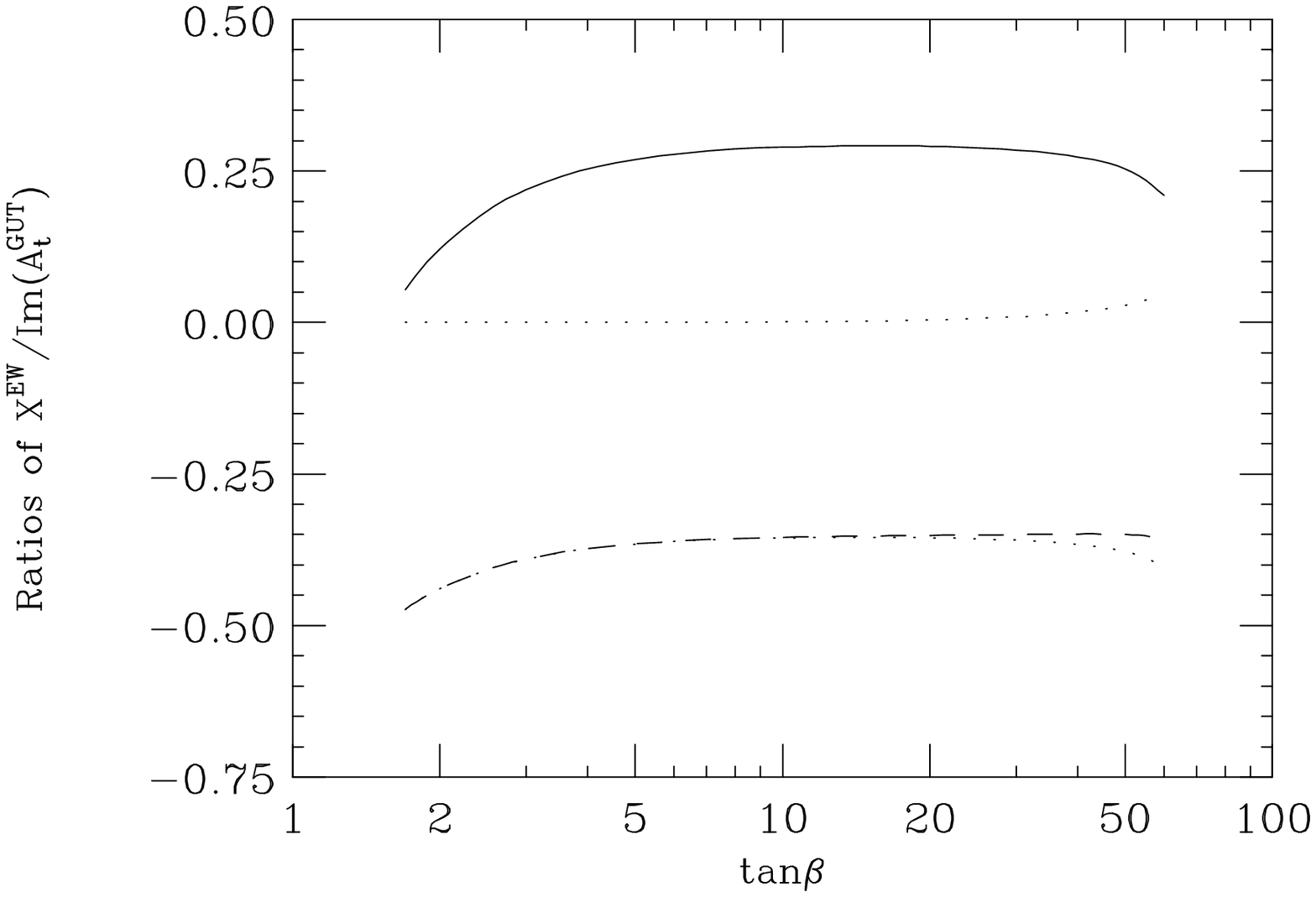}{The ratios of imaginary parts to $\Im\Agut_t$
versus $\tan\beta$ with $\Im\Agut_t \neq 0$ and $\Im\Agut_b=0$. 
The solid line is $\Im\Aew_t /\Im\Agut_t$; the
dashed line is $\Im\Bew /\Im\Agut_t$; and the upper (lower) dotted line
is $\Im\Aew_d /\Im\Agut_t$ ($\Im\Aew_u /\Im\Agut_t$).}

\jfig{atabfig}{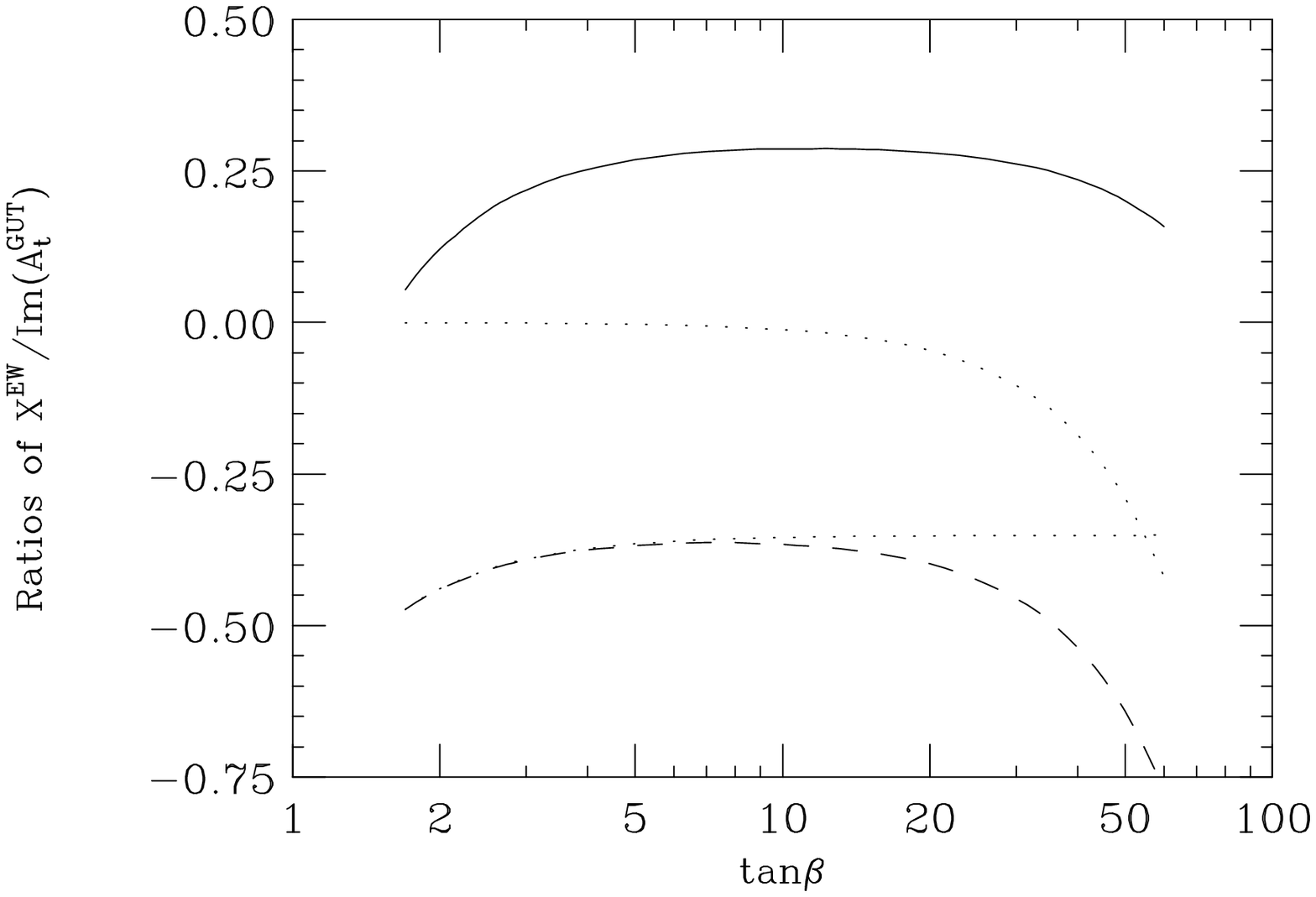}{Same as Fig.~2 with $\Im\Agut_b =\Im\Agut_t$.}

\jfig{atmabfig}{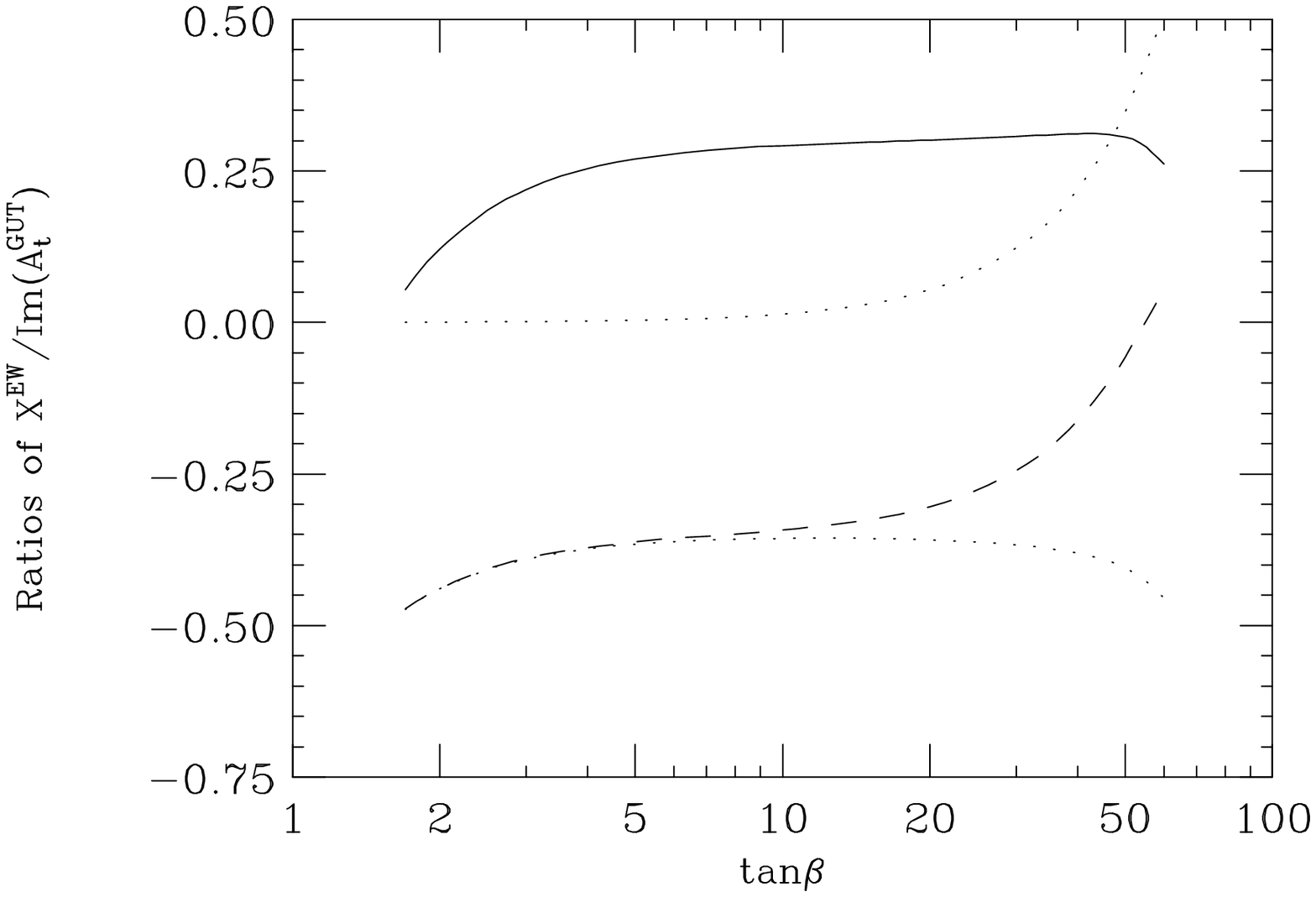}{Same as Fig.~2 with $\Im\Agut_b =-\Im\Agut_t$.}

\bfig{fig5}{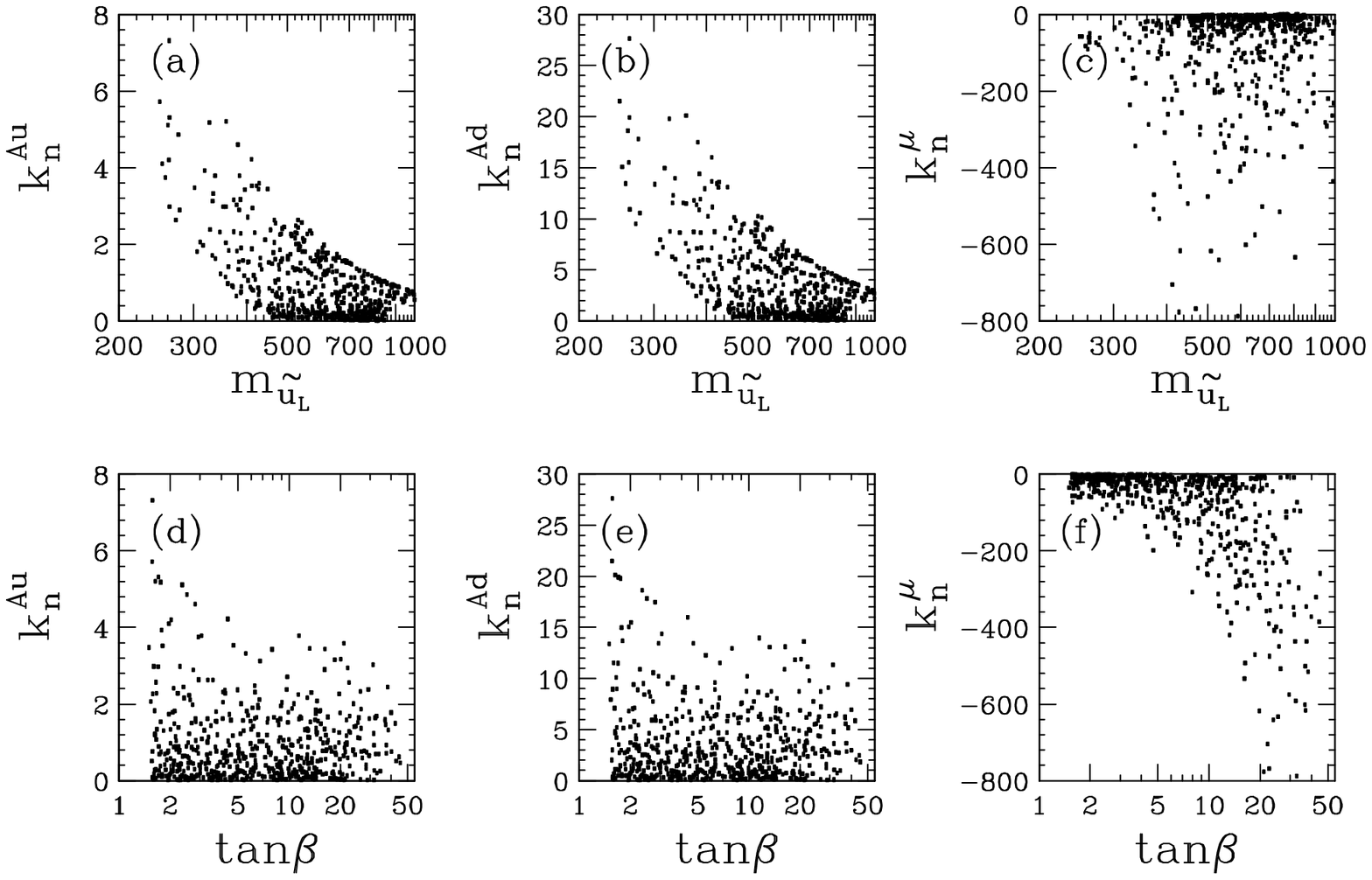}{Scatter plot of
(a and d) $k_n^{A_u}$,
(b and e) $k_n^{A_d}$, and
(c and f) $k_n^{\mu}$ versus squark mass and versus $\tan\beta$.
Each point represents a solution of the supersymmetric parameter space
with universal scalar and gaugino mass terms which is within
other experimental limits.} 

\bfig{fig6}{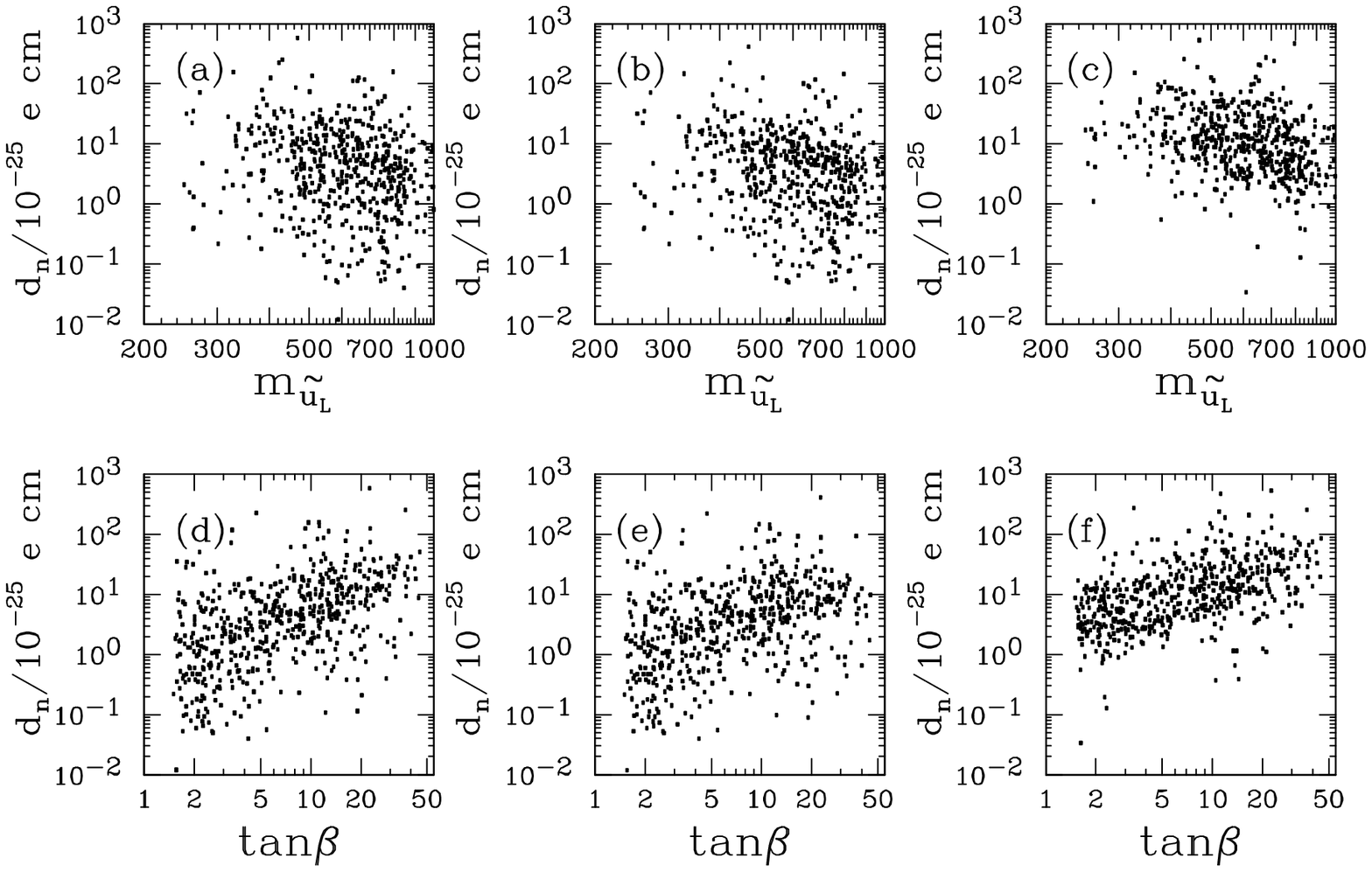}{Scatter plot of $d_n/10^{-25}e\cdot$cm 
versus squark mass and versus $\tan\beta$ for 
(a and d) $Arg A_t^{GUT}=Arg A_b^{GUT}=0.1$
with all other phases zero,
(b and e) $Arg A_t^{GUT}= -Arg A_b^{GUT}= 0.1$
with all other phases zero, and for
(c and f) universal phases $Arg A^{GUT} = Arg B^{GUT} = 0.1$.
Each point represents a solution of the supersymmetric parameter space
with universal scalar and gaugino mass terms which is within
other experimental limits.} 

\jfig{fig7}{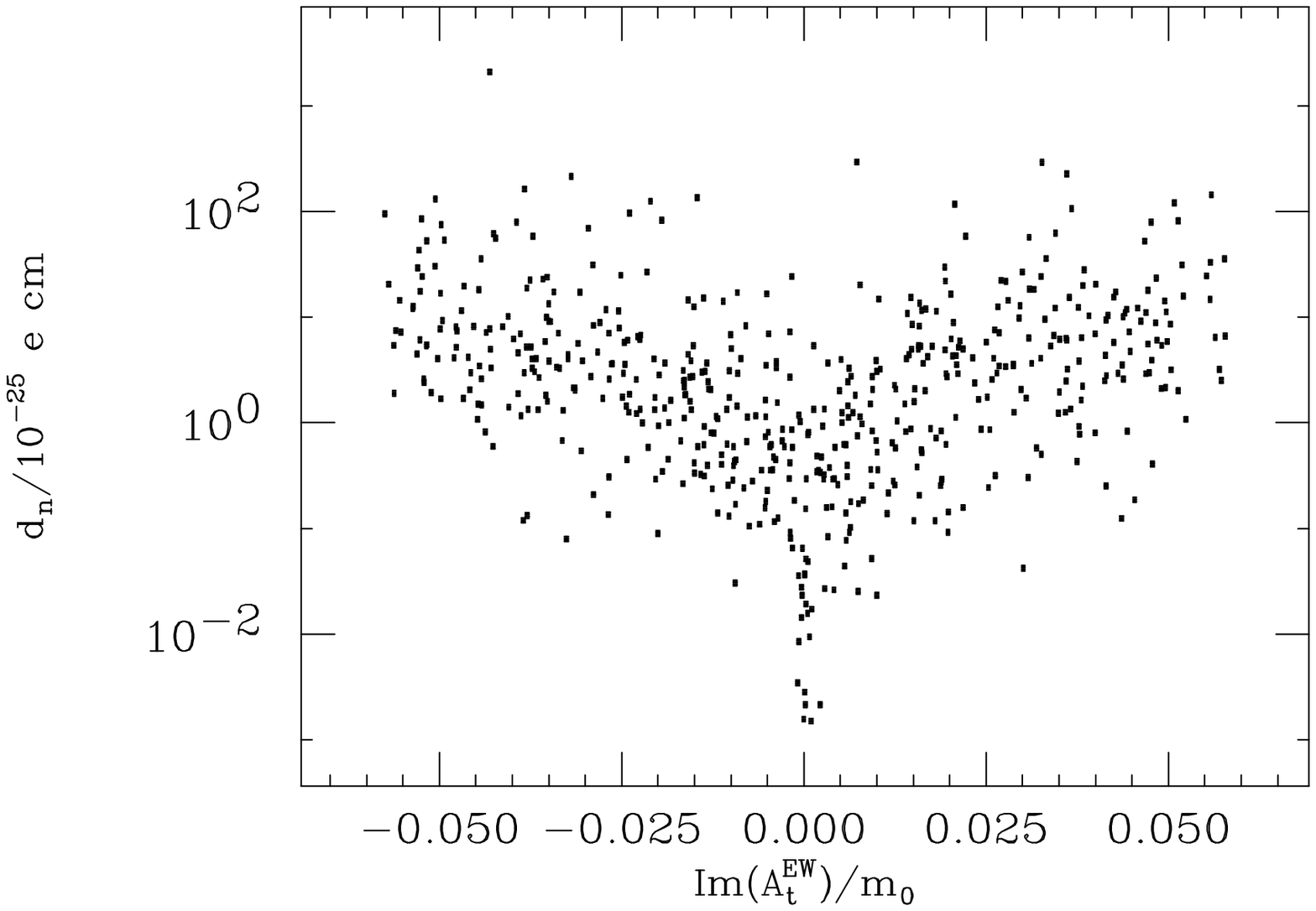}{Scatter plot of $d_n/10^{-25}e\cdot$cm
versus $Im A_t^{EW}/m_0$.  Each point represents a solution of the 
supersymmetric parameter space
with universal scalar and gaugino mass terms which is within
other experimental limits.} 


\end{document}